\documentclass[twocolumn,nofootinbib]{revtex4}
\usepackage{graphicx}
\usepackage{amsmath}

\begin{document}

\title{Beta relaxation in the shear mechanics of equilibrium viscous
  liquids: Phenomenology and network modeling of the alpha-beta merging region}

\author{Bo Jakobsen}
\email[Corresponding author. Email address: ]{boj@ruc.dk}
\author{Kristine Niss}
\author{Claudio Maggi}
\author{Niels Boye Olsen}
\author{Tage Christensen}
\author{Jeppe C. Dyre}

\affiliation{DNRF Centre ``Glass and Time'', IMFUFA, Department of Sciences, Roskilde University, Postbox 260, DK-4000 Roskilde, Denmark}

\begin{abstract}
  The phenomenology of the beta relaxation process in the
  shear-mechanical response of glass-forming liquids is summarized and
  compared to that of the dielectric beta process. Furthermore, we
  discuss how to model the observations by means of standard
  viscoelastic modeling elements. Necessary physical requirements to
  such a model are outlined, and it is argued that physically relevant
  models must be additive in the shear compliance of the alpha and
  beta parts. A model based on these considerations is proposed and
  fitted to data for Polyisobutylene 680.
\end{abstract}

\maketitle

\section{Introduction}
\label{}
Most knowledge about the beta (secondary) relaxation process in
glass-forming liquids has been obtained from dielectric
spectroscopy. This is due to the abundant amount of data available
from this technique, its high sensitivity and wide dynamical range
(routinely covering 9 decades). Dielectric spectroscopy, however, does
not probe the mechanical relaxation processes in the liquid. It is
therefore important to know to what extent the dielectric
phenomenology carries over to the mechanical relaxation functions. An
example where phenomenology does not carry over is the case of the
large Debye-like peak observed in the dielectric loss of monoalcohols,
which does not have any observable signature in the
frequency-dependent shear modulus \cite{Jakobsen2008}. There are also
cases where dielectric spectroscopy is not well suited for monitoring
the liquid's relaxations, e.g., if the liquid molecules have a very
small dipole moment (e.g., Squalane \cite{Richert2003a}) or if the
spectrum is dominated by other effects (e.g., dc conduction).

This motivates the interest in exploring the phenomenology of the beta
relaxation in other response functions. In the Roskilde group we have
developed a technique \cite{Christensen1995} for measuring the
frequency-dependent complex shear modulus,
$G(\omega)=G'(\omega)+iG''(\omega)$, which covers a rather large
frequency range compared to other mechanical spectroscopy techniques,
1mHz--10kHz, over moduli ranging from 0.1MPa to 10GPa. The technique
is based on piezoelectric ceramics acting as a converter from
  mechanical to electrical impedance. Our setup consists of a
piezoelectric shear-modulus gauge (PSG) in conjuntion with a cryostat
setup \cite{Igarashi2008a} and an electronics setup for measuring the
frequency-dependent capacitance \cite{Igarashi2008b}.

\begin{figure*}
  \centering 
 \includegraphics{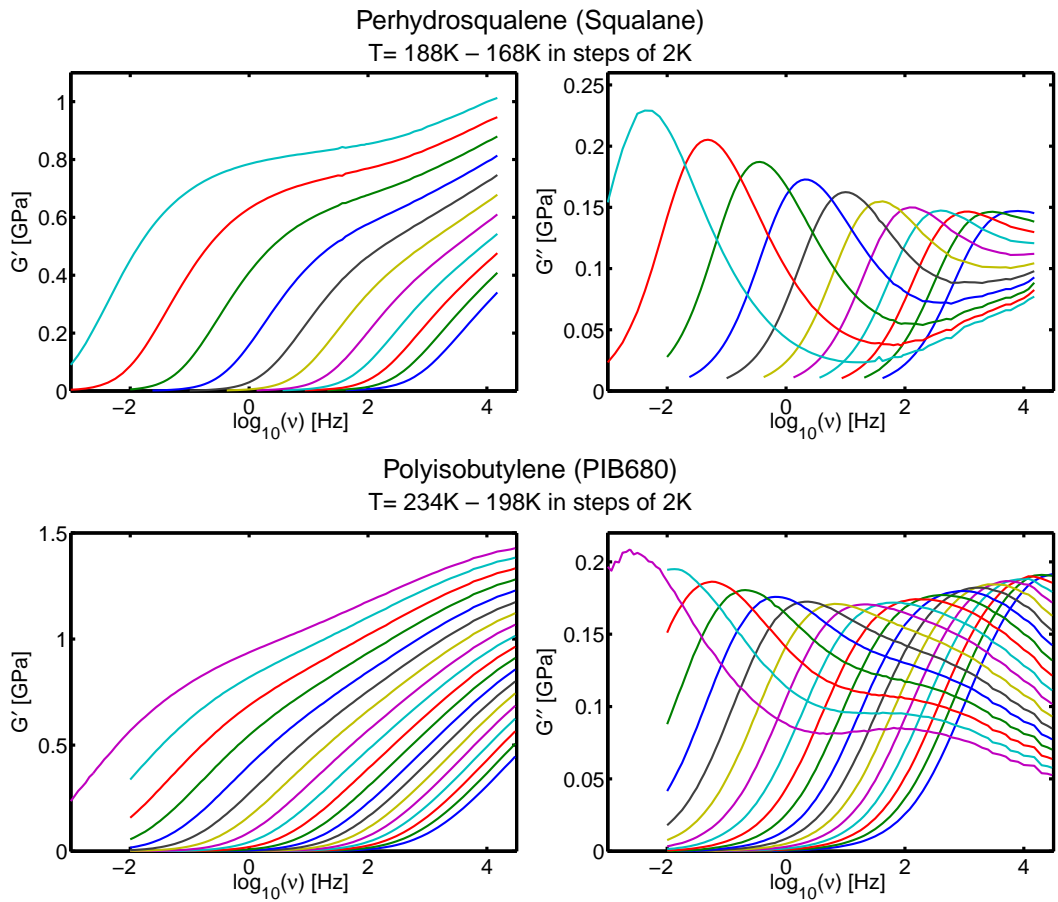}
 \caption{Shear-mechanical data on Perhydrosqualene (Squalane)
   (data originally presented in Ref.\ \cite{Jakobsen2005}) and
   Polyisobutylene (PIB680) (the Polyisobutylene sample has molecular
   weight $M_w=680\mathrm{g}/\mathrm{mol}$ and polydispersity
   $M_w/M_n=1.06$, the data originally appeared in Ref.\
   \cite{Niss2007PHD}).  The frequency-dependent complex shear
   modulus, $G(\nu)=G'(\nu)+iG''(\nu)$, is presented for each liquid
   for a number of iso-thermal frequency scans, showing real and
   imaginary parts.}
  \label{fig:Extreem}
\end{figure*}

With this technique the dynamical range is large enough to investigate
the beta relaxation in the shear mechanics. Figure \ref{fig:Extreem}
shows examples of shear-mechanical spectra in two cases where the beta
relaxation is very well resolved due to its position at low
frequency. We have in the past obtained shear-mechanical data on
several liquids, with and without a clearly resolvable beta 
relaxation\footnote{Data have been take on the following   substances (see the original publications for details):\\
  1,3-Butanediol \cite{Christensen1994}, 1,2,6-Hexanetriol
  \cite{Christensen1994c}, Tetramethyltetraphenyltrisiloxane (DC704)
  \cite{Christensen1994,Jakobsen2005,datarepro},
  2-Methyl-2,4-pentanediol \cite{Christensen1995}, Dibutylphthalate
  (DBP) \cite{datarepro,Behrens1996,Maggi2008}, Triphenylethylene
  (TPE) \cite{Jakobsen2005,datarepro}, Polyphenylether (PPE)
  \cite{Jakobsen2005,datarepro}, Perhydrosqualene (Squalane)
  \cite{Jakobsen2005,datarepro}, Polybutadiene
  \cite{Jakobsen2005,datarepro}, Decahydroisoquinoline (DHIQ)
  \cite{Jakobsen2005,datarepro}, Tripropyleneglycol (TPG)
  \cite{Jakobsen2005,datarepro}, Polyisobutylene (PIB680)
  \cite{Niss2007PHD} (this work \cite{datarepro}),
  Pentaphenyltrimethyltrisiloxane (DC705) \cite{datarepro,Maggi2008},
  1,2-Propanediol \cite{datarepro,Maggi2008}, Diethylphthalate (DEP)
  \cite{datarepro,Maggi2008}, m-Toluidine \cite{datarepro,Maggi2008},
  2-Butanol \cite{Jakobsen2008,datarepro}, 2-Ethyl-1-hexanol
  \cite{Jakobsen2008,datarepro}.\\
  Most of the data are available in electronic form from the ``Glass
  and Time: Data repository'' \cite{datarepro}, where figures off the
  available datasets can also be found for easy access.\\
  The Polyisobutylene (PIB680) data is on a sample with molecular
  weight $M_w=680\mathrm{g}/\mathrm{mol}$ and polydispersity
  $M_w/M_n=1.06$. The sample was acquired from Polymer Standart service,
  and used as received. Based on the shear mechanical data the glass
  transition temperature was found to be $T_g=195\mathrm{K}$, and the
  fragility index $m=80$ \cite{Niss2007PHD}.}; 
in most cases dielectric data have also
been taken (or exists in the literature). These measurements
constitute a database large enough for inquiring into the general
properties of the beta relaxation in shear mechanics and the relation
to the corresponding process in the dielectric response.

In section \ref{sec:Phenomenology} the general phenomenology of the
shear-mechanical beta relaxation is presented. Besides establishing
the phenomenology of the shear-mechanical beta relaxation, such data
can also be used to investigate different models. An important issue
is the way in which the alpha and beta relaxations merge and interact
in the region where their time scales are not well separated. Because
the shear modulus probes a fundamental property of the liquid, namely
its flow properties, constraints can be put on which models are
physically acceptable. In section \ref{sec:HowToMerge} we present such
considerations based on the class of simple element-based models. In
section \ref{sec:concr-fitt-model} a specific model is presented,
which is able to capture the full spectral shape during merging of the
alpha and beta processes, as shown by example for
Polyisobutylene. Section 5 gives a brief summary.

\section{Phenomenology of the shear-mechanical beta relaxation:
  Comparing to the dielectric beta relaxation}
\label{sec:Phenomenology}
The below phenomenology of shear-mechanical beta relaxations in
molecular liquids is based on a limited number of data sets, because
few techniques exist that can access the relevant frequency and
modulus ranges. Besides the above-mentioned measurements by our group,
we know of the following works comparing to the dielectric
phenomenology: The work on m-Toluidine by Mandanici \textit{et
al.} \cite{Mandanici2006} (see also Ref.\ \cite{Hutcheson2008}) and the
work on Ethylcyclohexane by Mandanici and Cutroni
\cite{Mandanici2007,Mandanici2009} (see  Ref.\ \cite{Mandanici2008}
for the dielectric data).

The below three points were in 2005 proposed by some of us based on
measurements on seven liquids \cite{Jakobsen2005}, and are largely
substantiated by the data sets existing today.  The entire
shear-mechanical beta relaxation peak is rarely visible within our
experimental frequency window, a fact that obviously complicates the
analysis quite a lot. We interpret the high-frequency rise, which is
sometimes seen in the imaginary part of the shear modulus (see e.g.,
the Squalane data in Fig.\ \ref{fig:Extreem}), as the signature of an
only partly visible beta relaxation process (peaking at a frequency
that is outside our measuring range). The observations are based on
this interpretation of the behavior of the high-frequency rise in the
imaginary part of the shear modulus.

\begin{figure*}
  \includegraphics{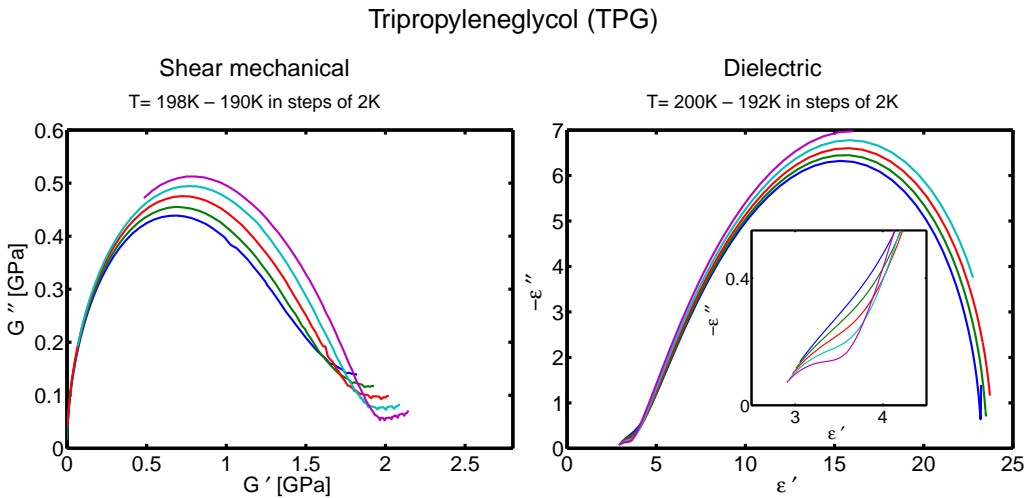}
  \caption{Shear-mechanical and dielectric data on  Tripropyleneglycol
    (TPG) (data originally presented in Ref.\ \cite{Jakobsen2005}). In
    each case a Nyquist plot (also known as a Cole-Cole plot) is
  shown, plotting the imaginary part against the real part. The insert
  shows a zoom-in of the dielectric beta relaxation.
  \label{fig:RelaxStrength}}
\end{figure*}

\begin{enumerate}
\item When a beta process is observed in the shear mechanics, it is also observed in dielectrics, and vice versa. \\
  In most cases this is a very clear observation; either a liquid has
  a beta process (e.g., Tripropyleneglycol on Fig.\
  \ref{fig:RelaxStrength}, Dibutylphthalate
  \cite{datarepro,Maggi2008,Albena09} and Polybutadiene
  \cite{Jakobsen2005,datarepro}), or it does not (e.g.,\
  Triphenylethylene and Tetramethyltetraphenyltrisiloxane (DC704)
  \cite{Jakobsen2005,datarepro}).

\item The shear-mechanical beta relaxation is shifted to higher frequencies compared to the dielectric beta relaxation. \\
  Interestingly, the same is true for the alpha relaxation, an
  experimental observation substantiated by many measurements from
  different groups, e.g.,
  \cite{Diaz-Calleja1993,Menon1994,Donth1996,Deegan1999,Schroter2000,Schroter2002}. The
  shear-mechanical alpha relaxation is normally shifted from 0.5 to 1
  decades up in frequency compared to the dielectric alpha relaxation
  at same temperature. m-Toluidine is an extreme case in this respect
  with a very small shift \cite{Mandanici2006,Hutcheson2008}.

\item The relaxation strength of the shear-mechanical beta relaxation, relative to the alpha relaxation, is larger than the corresponding ratio for the dielectric case. \\
  An extreme case is that of Tripropyleneglycol (Fig.\
  \ref{fig:RelaxStrength}), in which the dielectric beta relaxation is
  barely visible on the scale of the alpha relaxation, whereas the
  shear-mechanical beta relaxation is quite pronounced.
\end{enumerate}

These three points supplement the picture of the dielectric beta
  process detailed in three earlier publications of ours
\cite{Olsen1998a,Olsen2000,Dyre2003a}: The dielectric beta process is
well known to have a thermally activated loss-peak frequency in the
glass phase with an almost temperature-independent loss magnitude. In
the equilibrium liquid phase, however, where it is difficult to study
the dielectric beta process due to the merging with the alpha process,
the characteristics of the beta process are quite different
\cite{Olsen1998a,Olsen2000,Dyre2003a}. Here we find that the
dielectric loss peak frequency is either completely temperature
independent or only very weakly temperature dependent. In
  contrast, the beta loss peak magnitude is strongly temperature
  dependent in the liquid phase. In order to model these striking
features, as well as the annealing behavior of the beta process, in
2003 some of us proposed a ``minimal'' model \cite{Dyre2003a} for the beta
process based on an asymmetric potential with two minima. The two
energy parameters of this model were assumed to depend on one single
structural parameter, $s$, which freezes in the glass phase. This
model is able to account for many properties of the beta process, in
particular for the difference between liquid and glass behavior and
the beta annealing properties.

In the mechanical case we are unfortunately not able to study the beta
process in the glass phase with our PSG transducer, because the
calibration of the PSG parameters becomes uncertain at temperatures
deep below the glass transition of the investigated liquid. This is
due to the difference in the static stress between the filled and
empty transducer. It would certainly be worthwhile to compare the
mechanical beta relaxation in glass and liquid to see whether the
picture is similar to what is seen for dielectrics.

Currently no solid explanation exists for the differences between the
phenomenology of shear mechanical and dielectric alpha and beta
relaxations. The good news is that the shear-mechanical data confirm
that what we observe in dielectric spectroscopy is closely related to
the basic mechanical relaxations of the liquid, also on the level of
the beta process. The general shift between the time scales of the two
processes, which is in most cases virtually temperature independent,
is not surprising, given that such a shift often arises going
from one response function to another. A striking fact is the
generally observed enhancement of the relative relaxation strength of
the mechanical beta processes compared to that of the
dielectrics. This shows that conclusions based on dielectric
spectroscopy may lead to an underestimation of the importance of the
beta relaxation in the total relaxation picture.

\pagebreak
It is often argued that a connection between dielectric and
shear-mechanical relaxations must exist, with ideas dating back to
Debye's model of a rotating sphere, the rotation rate of which is
inversely proportional to the shear viscosity \cite{Debye1929}. It is
not the purpose of this paper to review or discuss such models in
detail, but it should be mentioned that some of us earlier argued
\cite {Christensen1994,Niss2005} that the Gemant-DiMarzio-Bishop model
\cite{Gemant1935,Dimarzio1974} captures the qualitative differences
between shear-mechanical and dielectric alpha and beta
relaxations. The model is a generalization of the Debye model
\cite{Debye1929} to the case of a frequency-dependent viscosity.
In Ref.\ \cite{Niss2005} we argued that the correct quantities to
compare are the shear modulus $G(\omega)$ and the \textit{rotational
dielectric modulus}, $\frac{1}{\epsilon(\omega)-n^2}$, where $n^2$
contains the contribution to the dielectric constant from electronic
polarization ($n^2=1+\chi_e$, with $\chi_e$
characterizing the electronic polarizability). This model is {\it
  qualitatively} consistent with points 1--3 above. The model,
however, does not fit data {\it quantitatively} (generally, it is not
possible to get the correct shape and position of the alpha peak of
the dielectric rotational modulus compared to the shear modulus (see
also Ref.\ \cite{Christensen1994}), the model furthermore predicts too large
a beta relaxation in the shear modulus).

Phenomenologically it is sometimes suggested that what should be
compared is the ``dielectric modulus'', $1/\epsilon(\omega)$, and the
shear modulus. However, in our opinion this is unphysical as one
trivially can write $\epsilon(\omega)=\chi_{r}(\omega)+n^2$, where
$\chi_r(\omega)$ is the contribution from rotation of the
molecules. $\chi_r(\omega)$ is the quantity that is assumed to be
related to the shear modulus/viscosity. The consequence is that
$\frac{1}{\epsilon(\omega)}=\frac{1}{\chi_{r}(\omega)+n^2}$, hence
that two liquids with the same $\chi_r(\omega)$ (and therefore same
molecular rotational dynamics) can have different dielectric modulus;
a related discussion took place in the 1990's in the ion conduction
field, see, e.g., \cite{Dyre1991,Ngai2000,Sidebottom2000}. The quantity
naturally arising from the Gemant-DiMarzio-Bishop model
\cite{Niss2005}, the rotational dielectric modulus,
$\frac{1}{\epsilon(\omega) - n^2}=\frac{1}{\chi_r}$, is on the
contrary only related to the rotational part of the dielectric
constant.

\section{Considerations on how to fit the full shear-mechanical spectrum}
\label{sec:HowToMerge}
There are several ways to model the phenomenology of the interplay
between the alpha and beta relaxation processes. A recent paper
\cite{Saglanmak2010} presents a new approach for the dielectric
alpha-beta merging and also briefly reviews other approaches. In the
present paper we present a related scheme for the alpha-beta merging
in shear-mechanical data. When formulating models for the alpha-beta
relaxation, the interplay of a number of physical constraints must be taken
into account. These constraints limit the number of possible models
for the full shear-mechanical spectrum, while there are no similar
constraints for modeling the dielectric spectrum.

The following analysis is limited to the classical class of
macroscopic viscoelastic models as described, e.g., by Harrison
\cite{Harrison1976}. The set of fundamental mechanical components
consists of 
\begin{eqnarray*}
  \label{eq:1}
  \begin{array}{lll}  
  1) &\text{an elastic spring:} & J(\omega)=1/G\\
  2) &\text{a dash pot:} & J(\omega)=\left(i\omega\eta\right)^{-1} \\
  3) &\text{a constant-phase element:} &
  J(\omega)=k\left(i\omega\right)^{-\alpha} \\ &&  \text{with}\ 0 \leq \alpha \leq 1,
  \end{array}
\end{eqnarray*}
where $J(\omega)$ is the complex frequency-dependent compliance
($J(\omega)=1/G(\omega)$). The elements are characterized by 1) a
spring constant ($G$), 2) the viscosity ($\eta$), and 3) a strength
($k$) and an exponent ($\alpha$). The constant phase element was
considered long ago by Cole and Cole in the dielectric context
\cite{Cole1941} and later by Jonscher
\cite{jonscher74,jonsch77,Jonscher1983}; it generalizes the
$\sqrt{i\omega\tau}$ element of the BEL model by Barlow et
al. \cite{Barlow1967b}. These elements can be connected in two
different ways, a mechanical series connection (with same stress on
each element and additive displacement, yielding additive shear
compliances) or a mechanical parallel connection (with same
displacement on each element and additive stress, implying additive
shear moduli)\footnote{The mechanical networks described herein can
  alternatively be expressed as electrical circuits by representing
  mechanical displacement as charge displacement and stress as voltage. The
  spring is then replaced by a capacitor and the dash pot by a
  resistor. Parallel and series connections are furthermore
  interchanged.}.

Combining such elements to model an observed response is today often
regarded as old-fashioned and providing little insight. We do see
advantages with this kind of phenomenological modeling, however; for
instance combining the elements will always lead to a model that is
consistent with basic requirements like causality, analyticity, and
positive dissipation. Possible connections between the alpha and
  beta relaxation (e.g.\ Refs.\ \cite{Olsen1998a,Olsen2000,Dyre2003a})
  may in some cases be integrated into this kind of models through
  parametric relations between the elements. However, as the models
  are intrinsic coarse grained, information and correlations on the
  more detailed level are lost (as e.g.\ information on correlations
  on the level of dynamical heterogeneity \cite{Bohmer2006}).  Of
course, this kind of modeling should only be regarded as a step
towards the ultimate, microscopic model, which will eventually provide
a complete description of the mechanical response of viscous liquids.

Physical constraints on the possible models come from the knowledge of
the limiting behavior of the shear-mechanical response functions:
\begin{eqnarray}
\label{eq:con1}
  \lim_{\omega\rightarrow\infty}G'(\omega)&=&G_{\infty}\\\
  \lim_{\omega\rightarrow 0 }G'(\omega)&=&0\\  
\label{eq:con3}
  \lim_{\omega\rightarrow 0}
  \eta(\omega)&=&\lim_{\omega\rightarrow 0 } G''(\omega)/\omega=\eta_0 
  < \infty.
\end{eqnarray}
The physical interpretation of these equations are as follows: (1) at
short times a liquid behaves as a pure spring, (2) at long times the
liquid's stresses relax to zero, (3) the dc limit of the viscosity
is a finite (real) constant.

\begin{figure}
  \centering
  \includegraphics{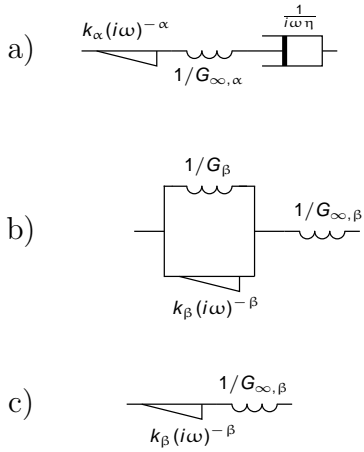}
  \caption{Mechanical network representation of the discussed
    individual model components. The building blocks are springs, dash
    pots, and constant-phase elements. The shear compliance is
    indicated for the individual elements.  \textbf{a)} Generalized
    BEL model used for the shear-mechanical alpha relaxation (modulus
    expressed in Eq.\ \eqref{eq:GenBell}). \textbf{b} and \textbf{c)}
    Cole-Cole like mechanical models used for modeling the mechanical
    beta relaxation (modulus expressed in Eqs.\ \eqref{eq:ColeCole}
    and \eqref{eq:ColeColeSimp} respectively).}
  \label{fig:AlphaAndBeta}
\end{figure}

Before discussing the merging of the alpha and beta relaxation we
need to choose models for the individual relaxations.  We found
that a generalized BEL model can fit the pure alpha relaxation very
well.  The generalized BEL model is illustrated in Fig.\
\ref{fig:AlphaAndBeta}a. It consists of a mechanical series connection
of a spring, a dash pot, and a constant-phase element (hence, it can
be seen as a generalization of the Maxwell model). The complex
frequency-dependent modulus is given by
\begin{eqnarray}
  \label{eq:GenBell}
G(\omega)=\frac{G_{\infty,\alpha}}{1+\frac{1}{i\omega\tau_{\alpha}}+q \left(\frac {1}
{i\omega\tau_\alpha}\right) ^{\alpha}} \,,
\end{eqnarray}
where $\tau_\alpha=\frac{\eta}{G_{\infty,\alpha}}$ (the Maxwell
relaxation time), and $q=G_{\infty,\alpha} k_\alpha
(\tau_\alpha)^\alpha$ in terms of the quantities defining the
individual model components.

It is well known that the dielectric beta relaxation can be fitted by
a Cole-Cole type function, which is symmetric in a log-log plot. From
the shear-mechanical data (e.g., Fig.\ \ref{fig:Extreem}) it is
observed that the shear-mechanical beta relaxation is also very broad
and has non-Debye low-frequency side. Since the information on the
high-frequency side of the mechanical beta relaxation is limited, we
assume that the mechanical beta process is also symmetrically
broadened; we thus fit it with a Cole-Cole type model.

There are two possible Cole-Cole type mechanical models giving a peak
in the modulus. Figures\ \ref{fig:AlphaAndBeta}b and
\ref{fig:AlphaAndBeta}c show these two models, which have shear
modulus given by, respectively,
\begin{align}
  G(\omega)&=\frac{G_{\infty,\beta}}{1+\frac{G_{\infty,\beta}/G_\beta}
    {1+(i\omega\tau_{\beta,J})^\beta}} & \ \text{for model (b)\ }\label{eq:ColeCole} \\
  G(\omega)&=\frac{G_{\infty,\beta}}{1+\left(\frac{1}{i\omega\tau_\beta}
    \right)^\beta} & \ \text{for model (c)}\,.\label{eq:ColeColeSimp}
\end{align}
Here the time scales are given as
$\tau_{\beta,J}=(G_{\beta}k_\beta)^{(-1/\beta)}$ and
$\tau_\beta=(G_{\infty,\beta}k_\beta)^{(-1/\beta)}$ in terms of the
quantities defining the individual model components.

\begin{figure}
  \centering
  \includegraphics{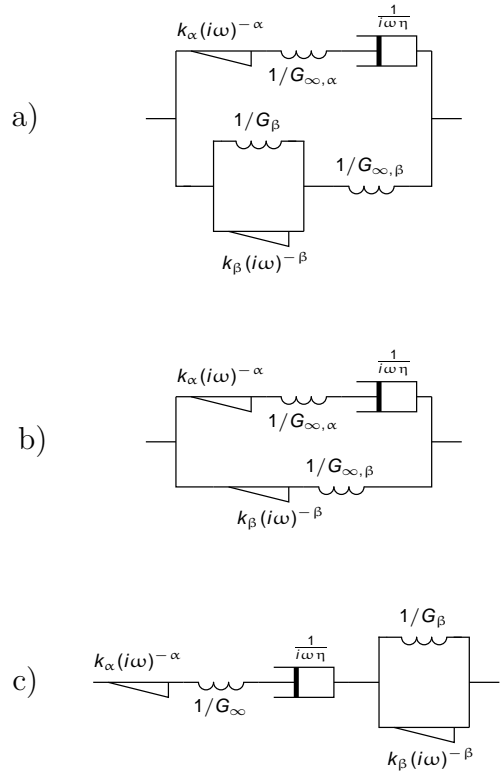}\\
  \caption{Mechanical network representation of the models for the full
    alpha and beta relaxation. The shear compliance is indicated for the
    individual  elements. \textbf{a}) and \textbf{b)} mechanical
    models which are additive in the shear modulus. \textbf{c)}
    mechanical model which is additive in the shear compliance 
    (modulus expressed in Eq.\ \eqref{eq:alpha-beta-shear}).}
  \label{fig:MergingModels}
\end{figure}

When constructing a combined model for the alpha and beta mechanical
relaxation, it is natural to try with an additive model in the measured
quantity, which in this case is the shear modulus. The two possible
additive models consisting of the generalized BEL model and one of the
Cole-Cole type models from Fig.\ \ref{fig:AlphaAndBeta} are shown in
Fig.\ \ref{fig:MergingModels}a and \ref{fig:MergingModels}b.

Model (a) has an obvious problem, namely that the modulus does not
decrease to zero in the low-frequency limit because the two springs of
the beta part will always be seen; likewise the model does not allow
for a dc flow. Model (b) also has problems. In this case the stress
does relax to zero at long times, but there is no finite dc viscosity
because of the constant-phase element. The extreme case $\beta=1$ is
an exception, but this corresponds to a beta relaxation with a Debye
shape, which is not observed. In conclusion, neither of the two
additive models comply with the physical constraints formulated in
Eqs.\ (\ref{eq:con1})--(\ref{eq:con3}).

Generally, if additivity is assumed in the shear modulus, the beta
element must allow for viscous flow in the low-frequency limit. This
unavoidably leads to Debye behavior at low frequencies, in
contradiction to what is observed. We therefore conclude that there
are no realistic models based on addition of the shear modulus of an
alpha and beta element.

The alternative is to use a model that is additive in the shear
compliance \cite{Ferry70}, i.e., is a mechanical series
combination of the basic elements. Figure \ref{fig:MergingModels}c
shows such a combination of the generalized BEL model and model (b)
from Fig.\ \ref{fig:AlphaAndBeta} (note that the two springs in direct
series have been merged into one,
$1/G_{\infty}=1/G_{\infty,\alpha}+1/G_{\infty,\beta}$). This model
has physically realistic properties, as it has an elastic behavior
at high frequencies and a viscous flow at low
frequencies.\footnote{An  model alternative, to the one discussed in
  detail, consists of the generalized BEL model in series with model
  (c) from Fig.\ 3. This does, however, not lead to a model function
  which resembles the alpha-beta merging phenomenology. The reason is
  that the model effectively only has one parameter determining both
  the time scale and strength of the beta relaxation part.}

The full expression for the shear modulus of our model, which obeys
the fundamental requirements, (Fig.\ \ref{fig:MergingModels}c) is
given by
 \begin{eqnarray}\label{eq:alpha-beta-shear}
G(\omega)=\frac{G_{\infty}}{{1+\frac {1}{i\omega\tau_\alpha}+q \left(\frac{1}{i\omega\tau_\alpha} \right)^{\alpha}}+{{\frac{G_{\infty}/G_{\beta}}{1+\left( i{\omega}\,
{{\tau}}_{{\beta,J}} \right) ^{\beta}}}} }.  
\end{eqnarray}
In the limit $\tau_\alpha \rightarrow \infty$ we regain the original
beta model (with $G_{\infty,\beta}=G_{\infty}$). If $\tau_{\beta,J} \rightarrow \infty$, we likewise
recover the original alpha model (with $G_{\infty,\alpha}=G_{\infty}$).

The phenomenology of this model is quite different from that of a
model that is additive in the shear moduli. The most significant
difference is that the beta relaxation does not show up on the
low-frequency side of the alpha relaxation if the beta process happens
to take place at lower frequencies than the alpha process.  In the
work of Sa\u{g}lanmak \emph{et al}. \cite{Saglanmak2010} a dielectric
model based on this shear-mechanical model is discussed in
detail\footnote{The network model for dielectric relaxation discussed
  in Ref.\ \cite{Saglanmak2010} may be regarded as a combination of
  this shear-mechanical model with the Gemant-DiMarzio-Bishop
  model. For a detailed analysis of the validity of the
  Gemant-DiMarzio-Bishop model see Ref.\ \cite{Niss2005}.}, and most
of the features can be directly transfered to this pure
shear-mechanical model.

Models capturing the alpha-beta merging are often used to investigate
the time scale and strength of the beta relaxation corrected for the
influence of the alpha relaxation. For such models it is often
possible to analytically determine the loss-peak frequency in both
the shear modulus and shear compliance of the pure beta model. For our
concrete model (the $\tau_\alpha \rightarrow \infty$ limit of Eq.\
\eqref{eq:alpha-beta-shear}, corresponding to Eq.\ \eqref{eq:ColeCole}
with the modified high-frequency spring) it is found that
 \begin{eqnarray}
  \omega_{\text{lp},\beta,J} &=& \frac{1}{\tau_{\beta,J}} \label{eq:lpj}\\
  \omega_{\text{lp},\beta,G} &=& \left(\frac{G_\infty+G_\beta}{G_\beta} \right)^{\frac{1}{\beta}} \frac{1}{\tau_{\beta,J}}\label{eq:lpg}.
\end{eqnarray}
The difference between the two time scales illustrates that the
observed time scale depends significantly on which response function
is probed. From the pure beta model it is further possible to find the
low-frequency limit of the shear modulus as
\begin{eqnarray}
  \label{eq:LowFreqLimit}
  G_{\beta,0}=\frac{G_\beta G_\infty}{G_\beta+G_\infty},
\end{eqnarray}
and hence the apparent beta relaxation strength $\Delta
G_\beta=G_\infty-G_{\beta,0}$.

\section{Fitting the new model for the combined shear-mechanical alpha-beta spectrum to data}
\label{sec:concr-fitt-model}
\begin{figure}
  \centering
    \includegraphics{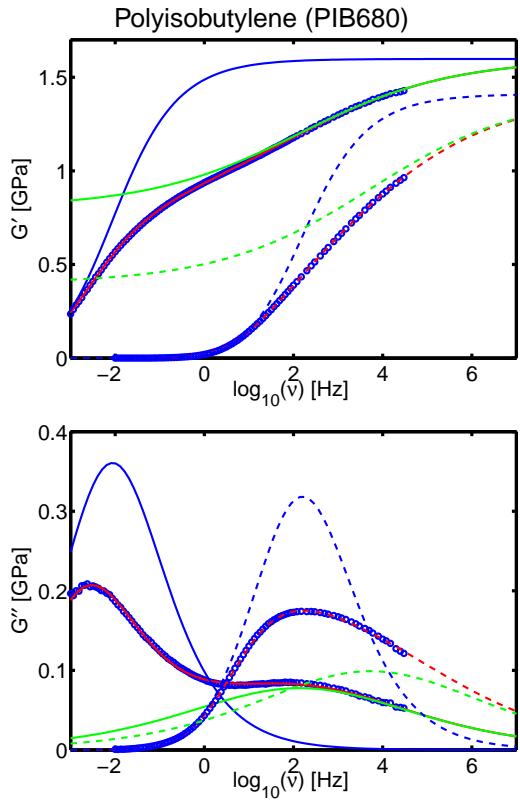}
    \caption{Fit of full alpha-beta model function, Eq.\
      \eqref{eq:alpha-beta-shear}, to data on Polyisobutylene (PIB680)
      at two temperatures, T=198K (full lines) and 216K (dashed
      lines); open circles are raw data. The high-frequency slope of
      the alpha relaxation was fixed to $-1/2$; the additional
      shape parameters of the alpha and beta relaxation are determined
      from the fit to the lowest temperature and then fixed for the
      fit at higher temperature (see section
      \ref{sec:concr-fitt-model} for details on the procedure).  Red
      lines are the fit of the full alpha-beta model Eq.\
      \eqref{eq:alpha-beta-shear}, blue lines the corresponding pure
      alpha part as given in Eq.\ \eqref{eq:GenBell} with
      $G_{\infty,\alpha}=G_\infty$, green lines are the corresponding
      pure beta part as given in Eq.\ \eqref{eq:ColeCole} with
      $G_{\infty,\beta}=G_\infty$.}
  \label{fig:Fit}
\end{figure}
\begin{figure}
  \centering
    \includegraphics{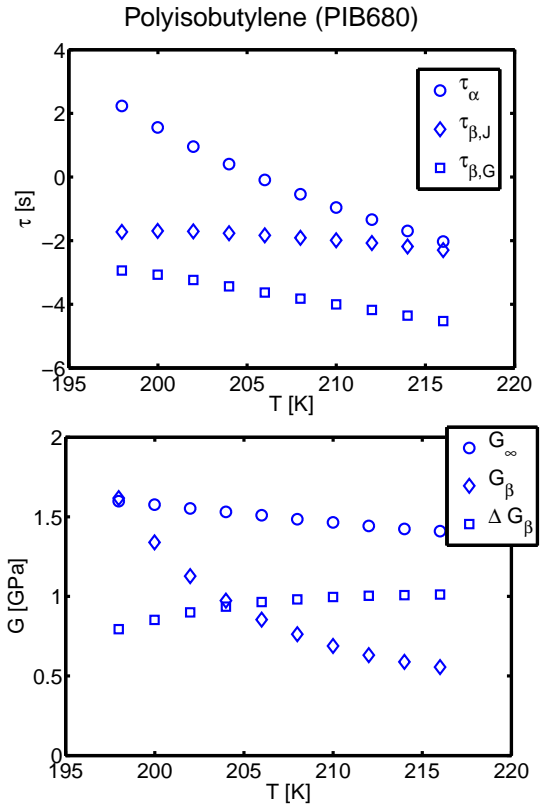}
    \caption{Resulting parameters from fitting the full alpha-beta
      model function, Eq.\ \eqref{eq:alpha-beta-shear}, to the
      shear-mechanical data on Polyisobutylene (PIB680) at the lowest
      ten temperatures (see section \ref{sec:concr-fitt-model} for
      details on the fitting procedure). \textbf{Top)} The
      characteristic alpha time scale, $\tau_{\alpha}$, and the two
      characteristic beta time scales, $\tau_{\beta,J}$ and
      $\tau_{\beta,G}$. \textbf{Bottom)} High-frequency elastic
      modulus, $G_\infty$, modulus of beta model spring, $G_\beta$,
      and the apparent beta relaxation strength $\Delta
      G_\beta=G_\infty-G_{\beta,0}$ (with $G_{\beta,0}$ given in Eq.\
      \eqref{eq:LowFreqLimit}).}
  \label{fig:Fitparm}
\end{figure}
To test the ability to fit data with the combined model as expressed
in Eq.\ (\ref{eq:alpha-beta-shear}) and illustrated in Fig.\
\ref{fig:MergingModels}c, the model was fitted to the ten lowest
temperatures of the Polyisobutylene (PIB680) data (Fig.\
\ref{fig:Extreem}). This furthermore illustrates what can be learned
from fitting a physically reasonable merging model to shear-mechanical
data.

The model has seven parameters, four related to the model for the pure
alpha relaxation and three to the beta model. This may seems
as a large number. However, it should be noted that an alpha model,
which has independent control over the high-frequency slope and the
width of the peak (or two other independent shape parameters, as in
the Havriliak-Negami equation), needs at least four
parameters. Likewise, a Cole-Cole type model needs three parameters
determining position, strength and shape.

The model can easily be fitted to spectra with well separated alpha
and beta relaxations (the lowest temperatures). When the alpha and
beta relaxations are close to merging, however, the limited frequency
range of the available shear-mechanical data makes this impossible.

Some of us have proposed that the generic alpha relaxation
process obeys time-temperature superposition (TTS) and has a
high-frequency slope of $-1/2$ \cite{Albena09,Olsen2001}. This was
originally based on dielectric data, but we later showed it to hold
reasonably well for shear-mechanical data as well
\cite{Jakobsen2005,Maggi2008}. The beta process is furthermore often
assumed to have a temperature-independent shape, as discussed above
(see also Ref.\ \cite{Olsen1998a}). 

Based on these observations the following fitting scheme was
applied. At the lowest temperature a least-squares fit was performed
with all parameters free except $\alpha=0.5$. Figure \ref{fig:Fit}
shows the result of this fit. The remaining temperatures were then
fitted with the shape-determining parameters, $q$ and $\beta$, fixed
to the found values. Figure \ref{fig:Fit} shows also the result of
these fits at a higher temperature.

Figure \ref{fig:Fitparm} shows $\tau_\alpha$, $\tau_{\beta,J}$, and
the time scale of the pure shear beta relaxation expressed as
$\tau_{\beta,G}=1/\omega_{\text{lp},\beta,G}$ (see Eq.\
  (\ref{eq:lpg})). The shear compliance time scale, $\tau_{\beta,J}$, is
almost temperature independent, while the shear modulus time scale,
$\tau_{\beta,G}$, is rather temperature dependent.  The elastic
parameters of the instantaneous modulus, $G_{\infty}$, and the beta
spring, $G_{\beta}$, are also reported in Fig.\ \ref{fig:Fitparm}, together
with the apparent beta relaxation strength.

The results shown here illustrate that hypothesis regarding spectral
shape and the temperature dependence of such can be tested with the
use of such an alpha-beta merging model. The present example shows
that within the framework of this merging model, data on
Polyisobutylene (PIB680) are consistent with the assumed
high-frequency slope of the alpha peak of $-1/2$ and a
temperature-independent shape of both the alpha and the beta
processes.

\section{Summary}
The shear-mechanical beta-relaxation phenomenology was investigated
based on data for eighteen liquids. The beta relaxation observed in
dielectric spectroscopy corresponds to that of shear-mechanical
spectroscopy and vice versa. Just as for the alpha
process, there is also a separation of the observed characteristic
time scales of the beta process in shear mechanics and dielectrics,
with the shear response being somewhat faster in both the alpha and beta
cases. The relaxation strength of the beta process relative
to the alpha relaxation strength is larger for shear than for
dielectric relaxation.

Within the class of viscoelastic models constructed from basic
elements a physically sound model must be additive in the shear
compliances, not in the shear moduli. A specific model was
constructed consisting of a generalized BEL model for the alpha
relaxation and a Cole-Cole type model for the beta. This model was
fitted to data on Polyisobutylene (PIB680); it was shown that good
fits can be obtained under the assumptions of a high-frequency slope
of $-1/2$ of the alpha relaxation and temperature independent shape
parameters for both the alpha and the beta relaxations. This
substantiates the conjecture that a $-1/2$ slope of the alpha loss (in
a log-log plot), as well as TTS, are generic features of the alpha
process \cite{Albena09,Olsen2001}.

\section*{Acknowledgments}
The centre for viscous liquid dynamics ``Glass and Time'' is sponsored
by the Danish National Research Foundation (DNRF).


\end{document}